\documentclass[
aps,
prb,
 amsmath,amssymb,
 superscriptaddress,
preprint,%
longbibliography,nobibnotes]{revtex4-1}

\usepackage{graphicx}
\usepackage{dcolumn}
\usepackage{bm}
\usepackage{ulem}
\usepackage[utf8]{inputenc}
\usepackage[T1]{fontenc}
\usepackage{mathptmx}
\usepackage{textcomp} 
\usepackage{nicefrac}
\usepackage{xcolor}
\usepackage{gensymb}
\usepackage{multibib}

\usepackage{lineno}

\usepackage{physics}
\usepackage{amsmath}

\DeclareMathAlphabet{\pazocal}{OMS}{zplm}{m}{n}
\newcommand{\Tb}{\pazocal{T}}

\newcommand{\Da}{\mathcal{D}}
\newcommand{\Ea}{\mathcal{E}}

\begin{document}

\title[Ultrafast nano-imaging of dark excitons]{Ultrafast nano-imaging of dark excitons}


\author{David Schmitt} %
\address{I. Physikalisches Institut, Georg-August-Universit\"at G\"ottingen, Friedrich-Hund-Platz 1, 37077 G\"ottingen, Germany}

\author{Jan Philipp Bange} %
\address{I. Physikalisches Institut, Georg-August-Universit\"at G\"ottingen, Friedrich-Hund-Platz 1, 37077 G\"ottingen, Germany}

\author{Wiebke Bennecke} %
\address{I. Physikalisches Institut, Georg-August-Universit\"at G\"ottingen, Friedrich-Hund-Platz 1, 37077 G\"ottingen, Germany}

\author{Giuseppe Meneghini} 
\address{Fachbereich Physik, Philipps-Universit{\"a}t, 35032 Marburg, Germany}

\author{AbdulAziz AlMutairi} 
\address{Department of Engineering, University of Cambridge, Cambridge CB3 0FA, U.K.}

\author{Marco Merboldt} %
\address{I. Physikalisches Institut, Georg-August-Universit\"at G\"ottingen, Friedrich-Hund-Platz 1, 37077 G\"ottingen, Germany}

\author{Jonas Pöhls} %
\address{I. Physikalisches Institut, Georg-August-Universit\"at G\"ottingen, Friedrich-Hund-Platz 1, 37077 G\"ottingen, Germany}

\author{Kenji Watanabe} %
\address{Research Center for Electronic and Optical Materials, National Institute for Materials Science, 1-1 Namiki, Tsukuba 305-0044, Japan}

\author{Takashi Taniguchi} %
\address{Research Center for Materials Nanoarchitectonics, National Institute for Materials Science,  1-1 Namiki, Tsukuba 305-0044, Japan}

\author{Sabine Steil} 
\address{I. Physikalisches Institut, Georg-August-Universit\"at G\"ottingen, Friedrich-Hund-Platz 1, 37077 G\"ottingen, Germany}

\author{Daniel Steil} %
\address{I. Physikalisches Institut, Georg-August-Universit\"at G\"ottingen, Friedrich-Hund-Platz 1, 37077 G\"ottingen, Germany}

\author{R. Thomas Weitz} %
\address{I. Physikalisches Institut, Georg-August-Universit\"at G\"ottingen, Friedrich-Hund-Platz 1, 37077 G\"ottingen, Germany}
\address{International Center for Advanced Studies of Energy Conversion (ICASEC), University of Göttingen, Göttingen, Germany}

\author{Stephan Hofmann} 
\address{Department of Engineering, University of Cambridge, Cambridge CB3 0FA, U.K.}

\author{Samuel Brem} 
\address{Fachbereich Physik, Philipps-Universit{\"a}t, 35032 Marburg, Germany}

\author{G.~S.~Matthijs~Jansen} %
\address{I. Physikalisches Institut, Georg-August-Universit\"at G\"ottingen, Friedrich-Hund-Platz 1, 37077 G\"ottingen, Germany}

\author{Ermin Malic} 
\address{Fachbereich Physik, Philipps-Universit{\"a}t, 35032 Marburg, Germany}

\author{Stefan Mathias} \email{smathias@uni-goettingen.de}%
\address{I. Physikalisches Institut, Georg-August-Universit\"at G\"ottingen, Friedrich-Hund-Platz 1, 37077 G\"ottingen, Germany}
\address{International Center for Advanced Studies of Energy Conversion (ICASEC), University of Göttingen, Göttingen, Germany}

\author{Marcel Reutzel} \email{marcel.reutzel@phys.uni-goettingen.de}%
\address{I. Physikalisches Institut, Georg-August-Universit\"at G\"ottingen, Friedrich-Hund-Platz 1, 37077 G\"ottingen, Germany}

\begin{abstract}

The role and impact of spatial heterogeneity in two-dimensional quantum materials represents one of the major research quests~\cite{Rhodes19natmat} regarding the future application of these materials in optoelectronics and quantum information science~\cite{Liu16natrevmats,Liang20advmat,mueller18}. In the case of transition-metal dichalcogenide heterostructures~\cite{Wang18rmp,Jin18natnano}, in particular, direct access to heterogeneities in the dark-exciton landscape~\cite{Malic18prm} with nanometer spatial and ultrafast time resolution is highly desired, but remains largely elusive~\cite{Plankl21natpho,Luo23nanolett}. Here, we introduce ultrafast dark field momentum microscopy to spatio-temporally resolve dark exciton formation dynamics in a twisted WSe$_2$/MoS$_2$ heterostructure with 55~femtosecond time- and 500~nm spatial resolution. This allows us to directly map spatial heterogeneity in the electronic and excitonic structure, and to correlate these with the dark exciton formation and relaxation dynamics. The benefits of simultaneous ultrafast nanoscale dark-field momentum microscopy and spectroscopy is groundbreaking for the present study, and opens the door to new types of experiments with unprecedented spectroscopic and spatiotemporal capabilities.

\vspace{2cm}

\end{abstract}

\maketitle


Direct nanoscale access to the exciton energy landscape and exciton dynamics is at the frontier of two-dimensional quantum materials research. Thinned down to the monolayer limit and stacked to form artificial heterosystems, these materials are highly tuneable and promise the realization of novel optoelectronic devices that can be used in applied~\cite{Liu16natrevmats,Liang20advmat,mueller18} and fundamental science~\cite{Novoselov12nat,Wang18rmp,Jin18natnano,Perea22apl}. The natural spatio-temporal regime for the quasiparticle dynamics in quantum materials is on the femtosecond-to-picosecond time and nanometer-to-micrometer spatial scale, respectively. In fact, this is the case for any optically-induced and laser-driven process in materials, reaching from direct laser-field manipulations of optically-induced phase transitions to novel transient states of matter~\cite{Basov17natmat,Torre21rmp}. For this reason, a multitude of large scale and laboratory-based research efforts currently aim to develop experiments that can directly access optoelectronics with spatio-temporal and possibly spectroscopic information~\cite{Cocker16nat,Garg20sci,Park16nanolett,dabrowski2020ultrafast,Plankl21natpho,Purz22jcp,Jakubczyk19acsnano,Danz21sci,Johnson23natphys,Luo23nanolett}. In these efforts, however, access to the spatio-temporal dynamics in two-dimensional semiconductors, e.g., transition metal dichalcogenides (TMDs), turns out to be particularly critical and challenging for two reasons: (i) Artificially stacked TMD heterostructures are known to exhibit a significant nanometer-scale heterogeneity~\cite{Rhodes19natmat} which represents a major research obstacle in TMD research with respect to future applications. (ii) The optoelectronic response of these materials is completely driven by Coulomb-correlated electron-hole pairs, i.e., excitons~\cite{Wang18rmp}. This strongly involves optically dark excitons~\cite{Malic18prm}, and therefore quasiparticles that are for the most part~\cite{Poellmann15natmat} inaccessible using current research approaches.

For the latter challenge, femtosecond momentum microscopy~\cite{medjanik_direct_2017,Keunecke20timeresolved}, a new variant of time- and angle-resolved photoelectron spectroscopy (trARPES), has just recently been shown to be exceptionally powerful to quantify the dynamics of bright and dark excitons in mono-~\cite{Madeo20sci,Wallauer21nanolett,Kunin23prl} and twisted bilayer~\cite{Schmitt22nat,Bange23arxiv,Karni22nat,Bange23arxiv2} TMDs. As a spatially-integrating method, however, typical trARPES experiments average over sample diameters of 10~µm or more, and are thus blind to nanoscale sample inhomogeneities~\cite{Haigh12natmat,Rhodes19natmat,Plankl21natpho,Purz22jcp}, dielectric disorder~\cite{Raja19natnano}, regions with strain gradients~\cite{Kumar15nanolett,Park16nanolett,Bai20natmat}, the presence of correlated phases in the nearest proximity~\cite{Xu20nat} and the local reconstructions of the moiré potential~\cite{Rosenberger20acsnano,deJong22natcom,Kapfer22arxiv}. In consequence, it is highly desirable to establish ultrafast ARPES on the nanoscale in order to image the energy landscape and dynamics of bright and dark excitons on femtosecond time- and nanometer length scales.

Here, we introduce ultrafast dark-field momentum microscopy as a new technique for simultaneous nano-imaging and nano-spectroscopy of ultrafast dynamics. We showcase the benefits of this technique by elucidating spatio-temporal and spatio-spectral dynamics of bright and dark excitons in a type-II band-aligned WSe$_2$/MoS$_2$ heterostructure. We find that this heterostructure, even in seemingly flat areas that exhibit high-quality photoemission spectra, shows distinct spatial heterogeneity. Specifically, we are able to map and correlate the variation of the energy landscape of excitons with its impact on the femtosecond interlayer exciton (ILX) formation dynamics. With many-particle modelling, we elucidate that this correlation is related to nanoscale variations of the strength of interlayer hybridization between the WSe$_2$ and the MoS$_2$ layers. Our study sets the stage for the future application of femtosecond dark-field momentum microscopy to the wide class of moiré heterostructures and low-dimensional quantum materials, studying optical excitations and light-induced emergent phenomena on the femtosecond time- and nanometer length scale.

\begin{figure}[]
    \centering
    \includegraphics[width=\linewidth]{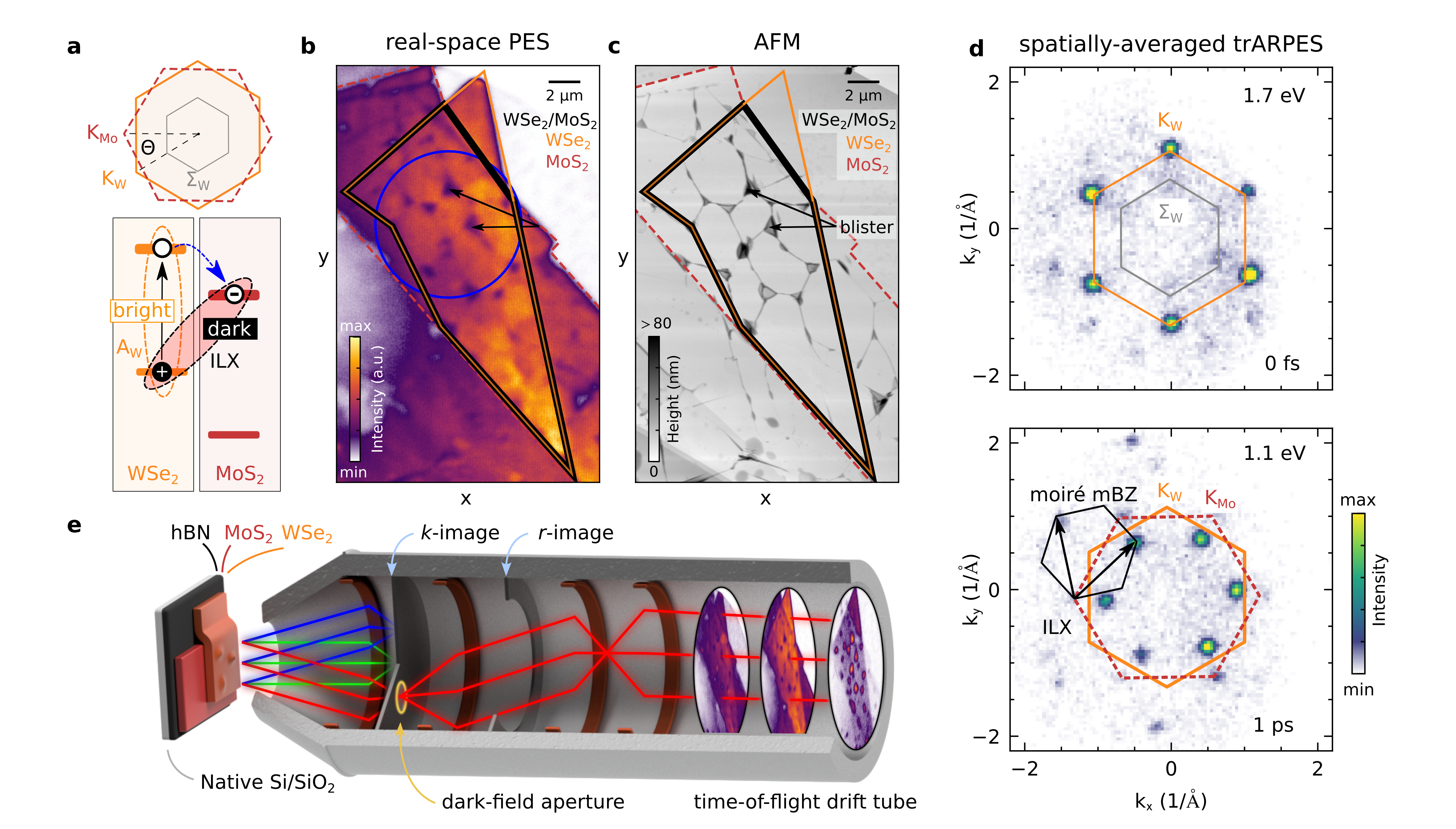}
    \caption{\textbf{Real- and momentum-space characterization of the WSe$_2$/MoS$_2$ heterostructure and working principle of the ultrafast dark-field momentum microscope.}
    \textbf{a} Single-particle picture of the type-II band aligned moiré heterostructure. The Brillouin zones of WSe$_2$ (orange) and MoS$_2$ (dark-red, dashed) are twisted by $\Theta=$28.8°$\pm$0.8°. Optically excited bright A$_{\rm W}$-excitons decay and form optically dark ILX.
    \textbf{b} Real-space resolved photoemission data showing the spatial alignment of the WSe$_2$ (orange) and MoS$_2$ (dashed dark-red) monolayers and their region of overlap (black). The black arrows label selected blisters.
    \textbf{c} AFM image of the same sample region.
    \textbf{d} Momentum-resolved photoemission maps from the intralayer A$_{\rm W}$-excitons, hybrid h$\Sigma$-excitons and the ILX measured in a spatially-averaged trARPES experiment with an aperture placed into the real-space plane of the momentum microscope (effective diameter of 10~$\mu$m, blue circle in \textbf{b}). Photoemisison yield is detected at the K$^{\left(\prime\right)}_{\rm W}$ valleys (orange hexagon), the $\Sigma_{\rm W}$ valleys (grey), the K$^{\left(\prime\right)}_{\rm Mo}$ valleys (dark red dashed) and the high-symmetry points of the moiré mini Brillouin zone (mBZ, black). Additional trARPES data obtained in the spatially averaged mode are shown in Extended Data Fig. 2.
    \textbf{e} Schematic illustration of the dark-field momentum microscopy experiment. In the electrostatic lens system of the microscope, $k$- and $r$-images are formed in the Fourier and the real-space plane, respectively. The kinetic energy of the photoelectrons is analyzed with a time-of-flight spectrometer. Dark-field momentum microscopy experiments can be performed by placing an aperture into the $k$-image to select a spectroscopic feature of interest (cf. red electron trajectories).
    }
\end{figure}

\vspace{1cm}

\noindent \textbf{Ultrafast dark-field momentum microscopy of nanoscale exciton dynamics}

We start our article with the lateral characterization of the 28.8°$\pm$0.8° twisted WSe$_2$/MoS$_2$ heterostructure. 
Using the real-space mode of the momentum microscope, Fig.~1b shows the spatial distribution of photoelectrons being emitted from the WSe$_2$/MoS$_2$ heterostructure (black polygon). In direct comparison with atomic force microscopy images (AFM, Fig.~1c and Extended Data Fig. 3), we find distinct spatial heterogeneity: Beside flat sample areas, significant topographical heights are found that can be attributed to residual gas and hydrocarbons trapped at the interface between both layers ('blisters', exemplary arrows in Fig.~1b,c). In this inhomogeneous heterostructure, the ultrafast exciton dynamics can be mapped by inserting an aperture into the real-space plane of the momentum microscope to select a region-of-interest  on the heterostructure flake (blue circle in Fig.~1b)~\cite{Schmitt22nat}. Already, this experimental scheme is extremely powerful in comparison to usual time-resolved ARPES experiments, because we can evaluate the pump-probe delay-dependent photoemission intensity from bright and dark excitons in a diameter of 10~µm. Figure~1d shows the respective momentum-fingerprints of optically excited A$_{\rm W}$-excitons (orange hexagon), the hybrid h$\Sigma$-excitons (grey) and the ILXs (black). By evaluating the pump-probe delay dependent photoemission yield from all excitons, consistent with our earlier work on a 9.8$\pm$0.8° twisted heterostructure~\cite{Schmitt22nat,Bange23arxiv2,Meneghini22naturalsciences}, we find that ILXs in the 28.8°$\pm$0.8° twisted heterosctructure are also formed in a two-step process via interlayer hybridized h$\Sigma$-excitons, i.e., via the cascade A$_{\rm W}\rightarrow$h$\Sigma\rightarrow$ILX (Extended Data Fig. 2). However, we emphasize that this analysis averages over all inhomogeneities within the region-of-interest. Specifically, the measurement neither excludes the contribution of blister areas, which clearly do not provide a well-defined WSe$_2$/MoS$_2$ interface, nor is it sensitive to possible heterogeneity of the energy landscape and dynamics of the excitons in seemingly smooth sample areas.




We overcome this limitation by combining our ultrafast approach with dark-field imaging in the momentum microscope~\cite{Barrett12rsi} to directly create nanoscale real-space snapshots of the femtosecond exciton dynamics (spatial resolution of 480$\pm$80~nm for the used microscope settings, cf. Extended Data Fig. 8). To do so, a circular aperture is placed in the Fourier plane of the microscope that blocks all photoelectrons except for those emitted in a specific circular momentum range, i.e., with the desired in-plane momenta k$_x$ and k$_y$, where excitonic photoemission signal emerges (cf. Fig.~1e; dark-field aperture indicated in inset of Fig.~2a,b, effective diameter of 0.4~\AA$^{-1}$). 
\begin{figure}[]
    \centering
    \includegraphics[width=.66\linewidth]{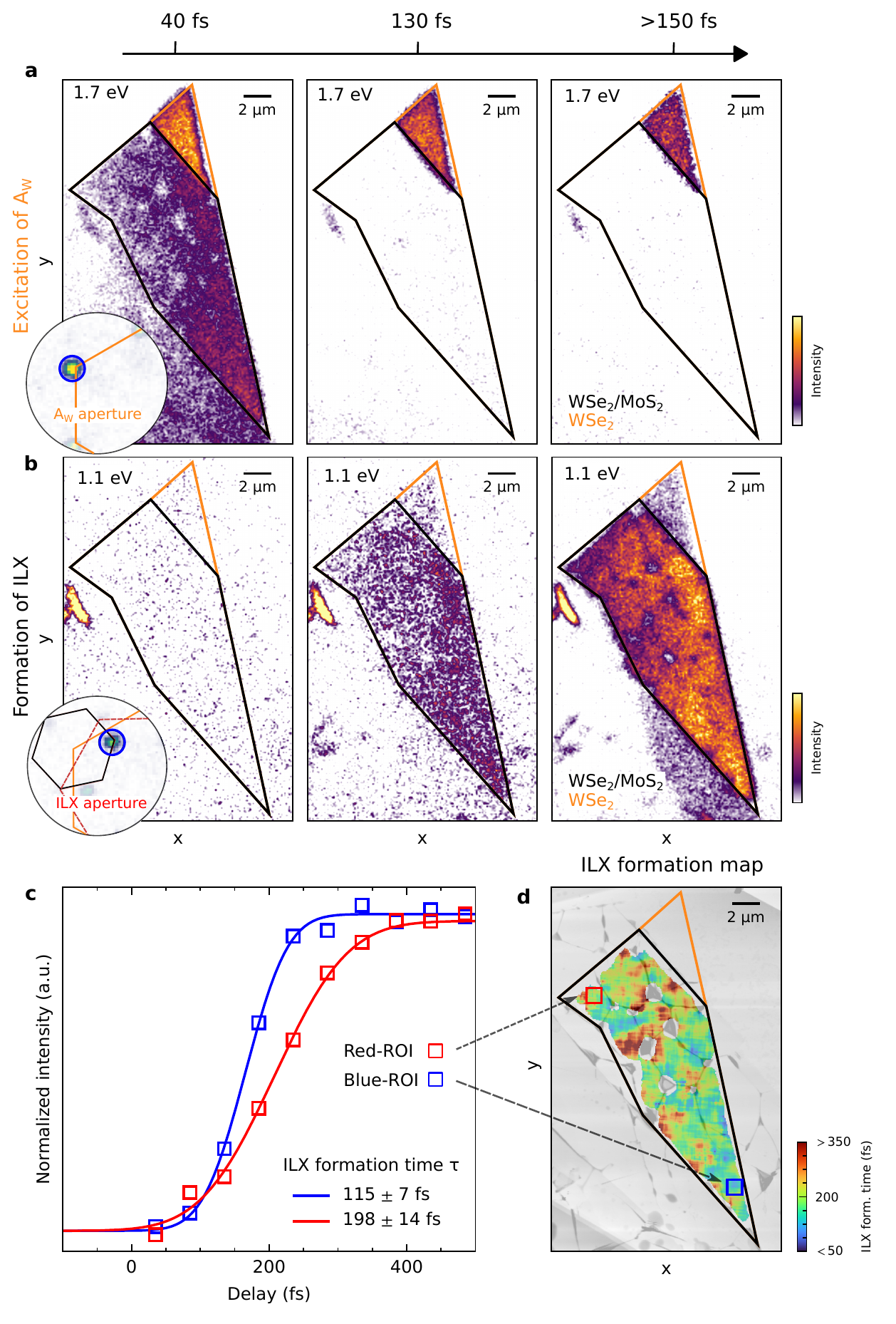}
    \caption{\textbf{Spatio-temporal snapshots of the formation and relaxation dynamics of bright A$_{\rm W}$-excitons and dark ILXs.}
    \textbf{a, b} Time- and real-space-resolved snapshots of the dark-field filtered photoemission yield from bright A$_{\rm W}$-excitons (\textbf{a}) and dark ILXs (\textbf{b}). The circular insets show the position of the dark-field aperture (blue circle) in the corresponding momentum-resolved measurement evaluated at the photoemission energies of the A$_{\rm W}$-exciton (1.7~eV above WSe$_2$ VBM) and the ILX (1.1~eV above WSe$_2$ VBM).
    \textbf{c} Pump-probe delay-dependent evolution of the ILX photoemission intensity evaluated in exemplary regions-of-interest (cf. red and blue squares in \textbf{d}).
    \textbf{d} The real-space dependence of the ILX formation time $\tau$ is color-coded on a heatmap. $\tau$ is only evaluated in regions where the superimposed AFM measurement (grey) shows heights <20~nm.
    }
\end{figure}
In this way, we are now able to create spatially-resolved snapshots of the photoelectrons being emitted from bright and dark excitons as a function of pump-probe delay (Supplemental Movies 1 and 2). Figure~2a shows the real-space formation and relaxation dynamics of the optically excited WSe$_2$ A$_{\rm W}$-exciton at 40~fs, 130~fs and >150~fs delay. Likewise, Fig.~2b shows the corresponding formation of the dark ILX. At first glance, it is apparent that the exciton dynamics differs in the WSe$_2$ mono- and the WSe$_2$/MoS$_2$ heterobilayer region: Photoemission yield from A$_{\rm W}$-excitons is found to be more intense in the WSe$_2$ monolayer than in the WSe$_2$/MoS$_2$ heterobilayer region. Notably, the A$_{\rm W}$-exciton photoemission intensity decays significantly faster in the heterobilayer region in comparison to the monolayer region, which is caused by the efficient relaxation channel into the hybrid h$\Sigma$-exciton (Fig.~2a, Extended Data Fig. 6, and ref.~\cite{Bange23arxiv2}). The ILX, on the other hand, are only formed in the region of overlap of the WSe$_2$ and MoS$_2$ monolayers. Moreover, it takes more than 200~fs until the photoemission yield of these excitons finally saturates (Fig.~2b and Extended Data Fig. 6).



Having recorded such spatio-temporally-resolved data of the bright and dark exciton densities, we can now quantitatively analyze the data for possible heterogeneity in the ILX formation time, which represents spatio-temporal experimental access to dark exciton dynamics that is so far unattainable with any other experimental method. Therefore, we evaluate the pump-probe delay-dependent evolution of the ILX photoemission intensity in a region of the size 1.2$\times$1.2~µm$^2$. While a certain degree of heterogeneity is to be be expected, we find that, astonishingly, the rise time of the ILX photoemission intensity can vary by more than a factor of 1.7: In the regions indicated by red and blue boxes in Fig.~2d, the ILX formation time changes from $\tau_{blue}=$115$\pm$7~fs to $\tau_{red}=$198$\pm$14~fs (Fig.~2c, error function based fit model described in methods). In order to map the nanoscale heterogeneities in the ILX formation time, we vary the selected region-of-interest pixelwise over the heterobilayer region and plot the formation time $\tau$ on a heat-map (Fig.~2d, relative fit error in Extended Data Fig. 7). Indeed, this heat-map shows a diverse distribution of the ILX formation time, with a gradient towards faster rise from left to right. Hence, although the spatially-averaged photoelectron spectra indicate a high-quality heterostructure (cf. Extended Data Fig. 5), the exciton dynamics are strongly affected. Importantly, because we have only evaluated areas that appear smooth in the AFM image (cf. AFM overlay and white areas in Fig.~2d), we exclude the possibility that blister-regions, where the interface between the WSe$_2$ and the MoS$_2$ layer is not well-defined, affect the observed dynamics.

Naturally, the question at hand is the microscopic origin of the spatially distinct ILX formation dynamics. Motivated by the experimental identification that the ILX is formed in a two-step process via layer-hybridized h$\Sigma$-excitons (cf. Extended Data Fig. 2), we make use of our recently developed microscopic model~\cite{Schmitt22nat,Meneghini22naturalsciences,Meneghini23}, and search for the key quantities that impact the ILX formation time in the exciton cascade A$_{\rm W}\rightarrow$h$\Sigma\rightarrow$ILX (Fig.~3a). In order to mimic a spatially inhomogeneous heterostructure with an inherently varying energy landscape~\cite{Raja19natnano,Cho18prb,Rhodes19natmat}, we systematically tune the relative energy alignment $\Delta E_{\rm sp}$ of the single-particle (sp) valence and conduction bands of WSe$_2$ and MoS$_2$, and solve the Wannier equation to calculate the energy landscape of the excitons. For increasing $\Delta E_{\rm sp}$, we find a systematic enhancement of the energies of the hybrid h$\Sigma$-exciton and the ILX (Fig.~3b), and, moreover, a changing degree of hybridization (DoH, c.f. Methods) of the h$\Sigma$-exciton (Fig.~3b). Next, within this energy landscape of excitons, we calculate the formation time of ILX by evaluating the dynamics within the excitonic density matrix formalism. The calculations shown in Fig.~3c predict that the formation time of the ILX decreases (increases) as the degree of hybridization of the h$\Sigma$-exciton is increased (decreased), consistent with earlier reports~\cite{Meneghini22naturalsciences,Merkl20natcom}. Most importantly, this prediction of the microscopic model has a significant implication for the interpretation of the femtosecond dark-field momentum microscopy data: First, the experimentally observed nanoscale variation of the ILX formation is an indication for a changing degree of interlayer hybridization between the WSe$_2$ and the MoS$_2$ layers. Second, the model calculations predict a correlation between the ILX formation time and the exciton energies of the h$\Sigma$-exciton and the ILX (Fig.~3c and 3d, insets). In the following, using the spatio-spectral mode of the dark-field momentum microscopy experiment, we aim to identify this correlation on the WSe$_2$/MoS$_2$ heterostructure.

\begin{figure}[]
    \centering
    \includegraphics[width=1\linewidth]{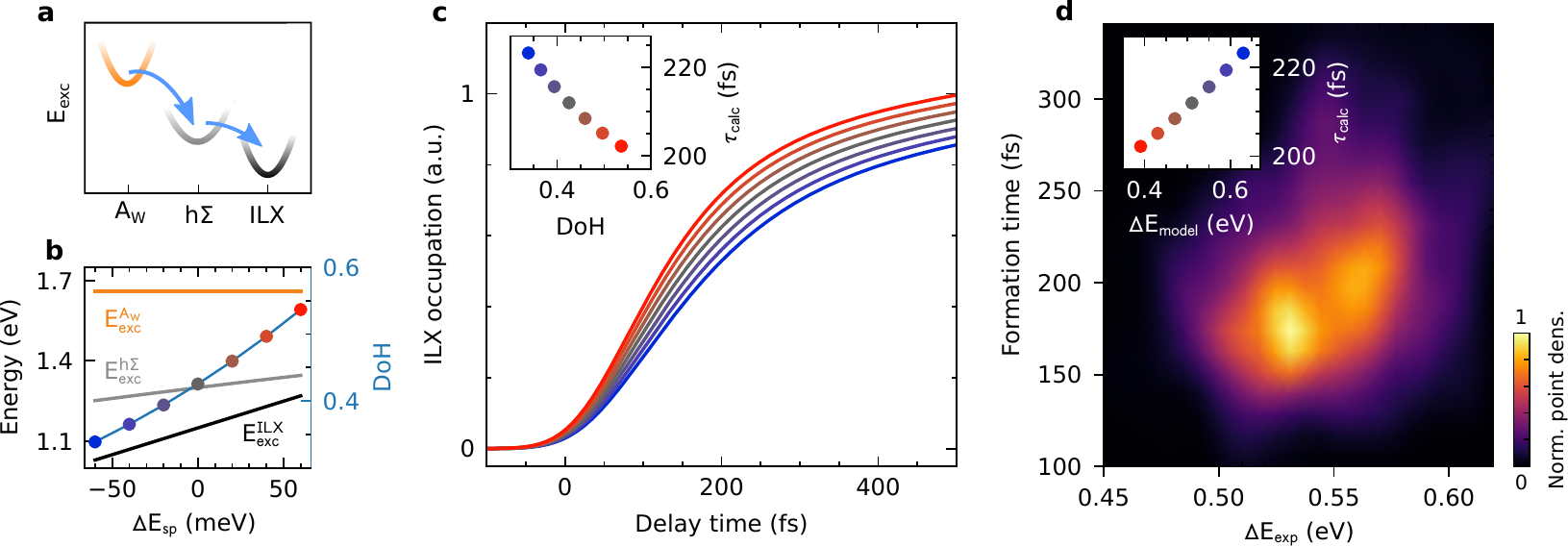}
    \caption{\textbf{ILX formation dynamics evaluated with many-particle calculations and correlation of the energy landscape of excitons and ILX formation dynamics.}
    \textbf{a} The mechanism of the ILX formation process occurs via the cascade A$_{\rm W}\rightarrow$h$\Sigma\rightarrow$ILX.
    \textbf{b} In the microscopic model, a spatially inhomogeneous WSe$_2$/MoS$_2$ heterostructure and a concomitant varying energy landscape of excitons is mimicked by the systematic variation of the relative band alignment $\Delta E_{\rm sp}$ of the single-particle bandstructures of WSe$_2$ and MoS$_2$ (cf. Fig.~1a). The exciton energies E$_{\rm exc}$ of the A$_{\rm W}$-exciton, the h$\Sigma$-exciton and the ILX are calculated as a function of $\Delta E_{\rm sp}$ by solving the Wannier equation. With increasing $\Delta E_{\rm sp}$, the degree of hybridization (DoH) of the h$\Sigma$-exciton (colored symbols) and the exciton energies of the h$\Sigma$-exciton (grey line) and the ILX (black line) increase.
    \textbf{c} The time-dependent occupation of ILX is calculated by solving the Heisenberg equation of motion for hybrid exciton densities, taking into account the $\Delta E_{\rm sp}$ dependent excitonic energies. An increasing degree of hybridisation of the h$\Sigma$-exciton leads to faster ILX formation times $\tau_{\rm calc}$ (inset).
    \textbf{d} We observe a correlation between the measured ILX formation time $\tau$ and the energy difference $\Delta E_{\rm exp} = E_{\rm exc}^{\rm A_W}-E_{\rm exc}^{\rm ILX}$ (main panel). The theoretical calculation, based on the assumption of a spatially dependent bandstructure alignment, predicts a decreasing formation time $\tau_{\rm calc}$ of ILX related to an increase of the hybrid character of the h$\Sigma$-exciton. A change in hybridization can be directly related to the quantity  $\Delta E=E_{\rm exc}^{\rm A_W}-E_{\rm exc}^{\rm ILX}$, that can be used in an experiment-theory comparison showing the same qualitative behaviour (inset).
    }
\end{figure}


\vspace{3cm}
\noindent \textbf{Spatio-spectro-temporal imaging of exciton dynamics}

\begin{figure}[]
    \centering
    \includegraphics[width=1\linewidth]{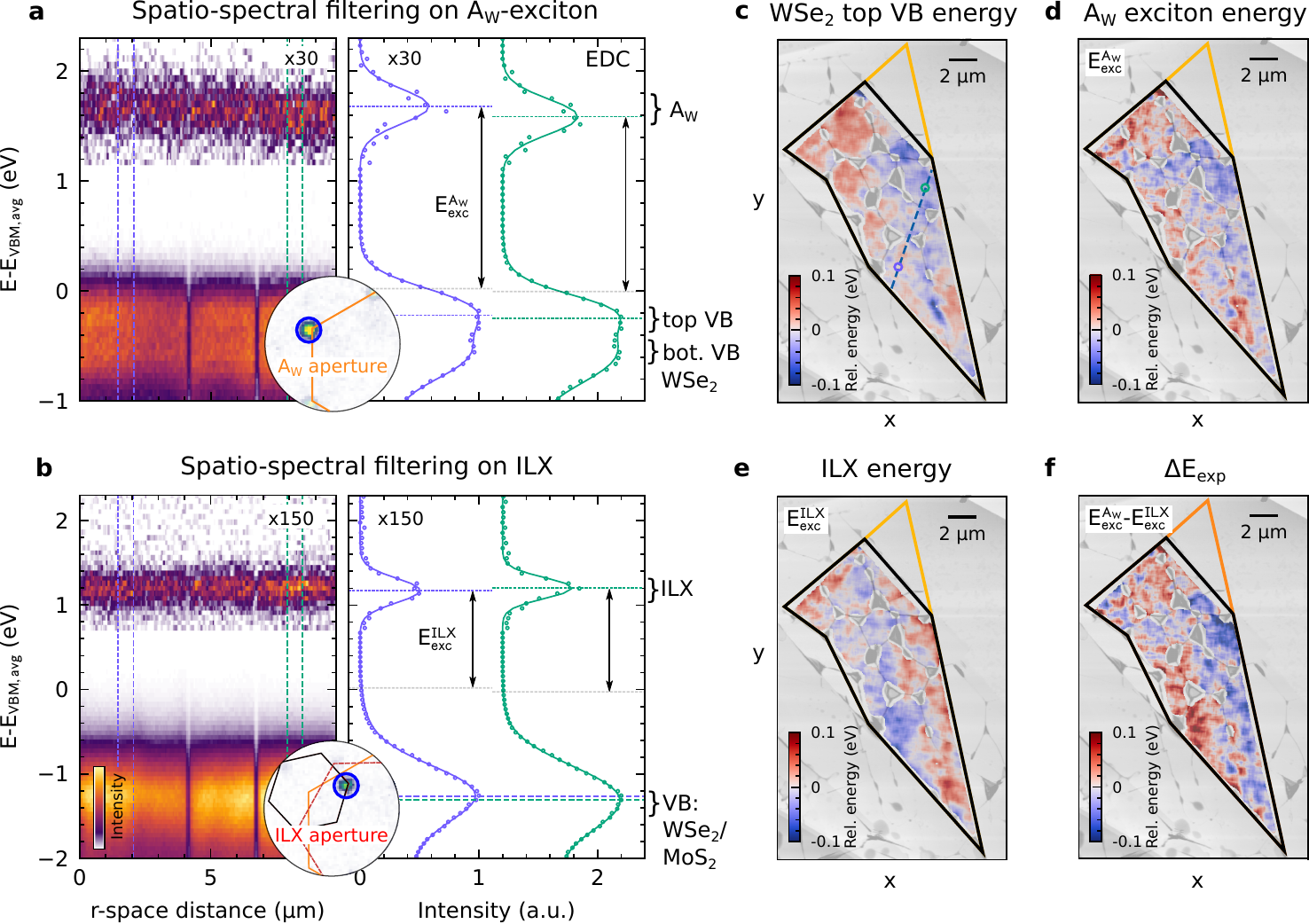}
    \caption{\textbf{Spatio-spectral imaging of the energy landscape of bright A$_{\rm W}$-excitons and dark ILXs.}
    \textbf{a},\textbf{b} Dark-field filtered photoemission intensity plotted as a function of photoelectron energy and the real-space distance along the blue dashed line in \textbf{c}. The data is generated by integrating over all measured pump-probe delays.
    \textbf{a} If the dark-field aperture is positioned on the momenta of the bright A$_{\rm W}$-exciton (inset), photoemission yield peaks at the energy of the A$_{\rm W}$-exciton (E-E$_{\rm VBM,avg}\approx1.7$~eV) and the spin-split valence bands of WSe$_2$ (E-E$_{\rm VBM,avg}\leq0$~eV). The spatio-spectral variation of the A$_{\rm W}$-exciton energy $E_{\rm exc}^{\rm A_W}$ can be read-out from such data by calculating the energy difference between the excitonic photoemission signal and the top WSe$_2$ valence band (cf. black vertical arrows in the EDCs and Extended Data Fig. 5). Note that because of the finite width of the dark-field aperture, the energy of the top WSe$_2$ valence band maxima needs to be corrected by 0.23~eV (cf. grey horizontal line centered at 0~eV and Extended Data Fig. 5).
    \textbf{b} If the same analysis is performed for the case that the dark-field aperture is positioned on the momenta of the ILX (inset), the spatio-spectral evolution of the exciton energy $E_{\rm exc}^{\rm ILX}$ can be extracted.
    \textbf{c-f} Two-dimensional heatmaps depicting the spatio-spectral variation of the \textbf{c} WSe$_2$ VBM energy, the \textbf{d} A$_{\rm W}$-exciton energy $E_{\rm exc}^{\rm A_W}$, the \textbf{e} ILX energy $E_{\rm exc}^{\rm ILX}$ and the \textbf{f} energy difference $\Delta E_{\rm exp} = E_{\rm exc}^{\rm A_W}-E_{\rm exc}^{\rm ILX}$. The energies are evaluated in sample regions with topographical heights <20~nm (cf. superimposed AFM image), and given with respect to their mean value.
    }
\end{figure}

The dark-field aperture inserted in the Fourier plane of the momentum microscope filters a momentum region of 0.4~\AA$^{-1}$ and blocks all remaining photoelectrons. Importantly, it transmits photoelectrons with all accessible kinetic energies within the selected momentum region that can then be analyzed with a time-of-flight spectrometer (cf. Fig.~1e). In consequence, we can not only monitor the ILX formation time, but also obtain direct spatio-spectroscopic insight to the energy landscape of bright and dark excitons. Figures~4a,b show such a dark-field-filtered data set that relates the energy-resolved photoemission spectra [i.e., energy-distribution-curves (EDCs)] with the real-space coordinate along the direction of the dashed line in Fig.~4c (resolution: 480$\pm$80~nm). If the dark-field aperture is positioned at the in-plane momentum of the A$_{\rm W}$-exciton (Fig.~4a), the data contains spatio-spectral information of the top and bottom valence bands (VB) of WSe$_2$ and the A$_{\rm W}$-exciton photoemission signal. Intriguingly, even in topographically smooth areas, we find inhomogeneous shifts of the photoemission signatures from the occupied valence bands and the A$_{\rm W}$-exciton (EDCs in Fig.~4a). For example, by plotting the energetic position of the top WSe$_2$ VB on a heat map, a nanoscale variation of the VB energy by $\pm$0.1~eV can be observed (Fig.~4c, resolution: 590~nm). We interpret this heterogeneous rigid energy shift as a direct indication for a nanoscale variation of the heterostructure homogeneity that can be attributed to, for example, dielectric disorder~\cite{Waldecker19prl,Raja19natnano}.

Even more interesting in terms of the optoelectronic response of the sample, our analysis provides unprecedented access to the energy landscape of bright \textit{and} dark excitons: The exciton energies E$_{\rm exc}^{\rm A_W}$ and E$_{\rm exc}^{\rm ILX}$ of the bright A$_{\rm W}$-exciton and the dark ILX can be quantified from the energy difference of the WSe$_2$ valence band maximum and the respective exciton photoemission signal~\cite{Weinelt04prl,Bange23arxiv,Bennecke23arxiv} (vertical arrows in Fig.~4a,b and Methods). In this manner, in Fig.~4d and 4e, the nanoscale variation of the energy of the A$_{\rm W}$-exciton and the ILX are visualized on a heat map (spatial resolution: 590~nm; error maps in Extended Data Fig. 7). Notably, we find that the exciton energies vary by $\pm$0.1~eV with a strong gradient from the left-hand-side to the right-hand-side of the heterostructure (Fig.~4d,e). 


\vspace{1cm}
\noindent \textbf{Correlating the spatio-temporal dynamics with the local exciton energy landscape}


Finally, we want to show how the measurement approach of the nanoscale spatio-temporal and spatio-spectral exciton dynamics  can be used to verify the predictions of our theoretical modelling. To do so, we generate a correlation map between the nanoscale ILX formation time and exciton energy landscape (Fig.~3d, main panel). While it would be most straightforward to correlate the energy of the hybrid h$\Sigma$-exciton with the ILX formation time, unfortunately, its photoemission intensity is too weak to be evaluated in a dark-field momentum microscopy experiment (Extended Data Fig. 2 and Methods). However, we can also focus on the energy separation $\Delta E_{\rm exp}$ between the optically-excited A$_{\rm W}$-exciton and the ILX (i.e., $\Delta E_{\rm exp} = E_{\rm exc}^{\rm A_W}-E_{\rm exc}^{\rm ILX}$, cf. Fig.~4f). Similar as for the hybrid h$\Sigma$-exciton, we expect that an increasing energy separation $\Delta E_{\rm model}=E_{\rm exc,\,model}^{\rm A_W}-E_{\rm exc,\,model}^{\rm ILX}$ leads to a slower formation time of the ILX (Fig.~3d, inset). Indeed, the experimental correlation map of these two quantities fully confirms the expectation that the ILX formation time becomes faster with decreasing $\Delta E_{\rm exp}$ (Fig.~3d). Hence, in qualitative agreement between experiment and theory, our analysis demonstrates that the nanoscale heterogeneitiy of the moiré heterostructure leads to distinct ILX formation times, which is dominantly caused by a changing degree of interlayer hybridization between the WSe$_2$ and the MoS$_2$ layers.


\vspace{1cm}
\noindent \textbf{Discussion}

Our work showcases how the femtosecond time-resolved realization of dark-field momentum microscopy enables access to a multitude of spectroscopic signatures with spatio-temporal and spatio-spectral resolution. In our study on a twisted WSe$_2$/MoS$_2$ heterostructure, we elucidate that the nanoscale heterogeneity of the heterostructure strongly affects the interlayer hybridization between the WSe$_2$ and the MoS$_2$ monolayers, and, accordingly, leads to a renormalization of the energy landscape of excitons and the formation time of ILX. In the future, ultrafast dark-field momentum microscopy will allow direct spatio-spectral and spatio-temporal access to, e.g., diffusion processes~\cite{Choi20sciadv,Rosati21natcom}, new correlated states of matter~\cite{Xu20nat}, and the real-space evolution of optically-induced phase transitions~\cite{Danz21sci}. More generally, it opens up unprecedented access to ultrafast quasiparticle dynamics in quantum materials on their fundamental time- and length scales.


\section{ACKNOWLEDGEMENTS}

This work was funded by the Deutsche Forschungsgemeinschaft (DFG, German Research Foundation) - 217133147/SFB 1073, projects B07 and B10, 432680300/SFB 1456, project B01 and 223848855/SFB 1083, project B9. A.A. and S.H. acknowledge funding from EPSRC (EP/T001038/1, EP/P005152/1). A.A. acknowledges financial support by the Saudi Arabian Ministry of Higher Education. E. M. acknowledges support from the European Unions Horizon 2020 research and innovation program under grant agreement no. 881603 (Graphene Flagship). K.W. and T.T. acknowledge support from the JSPS KAKENHI (Grant Numbers 20H00354, 21H05233 and 23H02052) and World Premier International Research Center Initiative (WPI), MEXT, Japan.

\section{AUTHOR CONTRIBUTIONS}
S.S., D.St., R.T.W., S.H., S.B., G.S.M.J., E.M., S.M. and M.R. conceived the research. D.Sch., J.P.B., W.B. and M.M. carried out the time-resolved momentum microscopy experiments. D.Sch. and J.P.B. analyzed the data. G.M. performed the microscopic model calculations. A.A. fabricated the samples. J.P. and D.Sch. performed the AFM measurements. All authors discussed the results. S.M. and M.R. were responsible for the overall project direction and wrote the manuscript with contributions from all co-authors. K.W. and T.T. synthesized the hBN crystals.



\clearpage
\section*{Methods}

\renewcommand{\figurename}{Extended Data Fig.}
\setcounter{figure}{0}


\vspace{.2cm}


\section{Heterostructure fabrication and twist-angle characterization}

\begin{figure}[htb!]
    \centering
    \includegraphics[width=\linewidth]{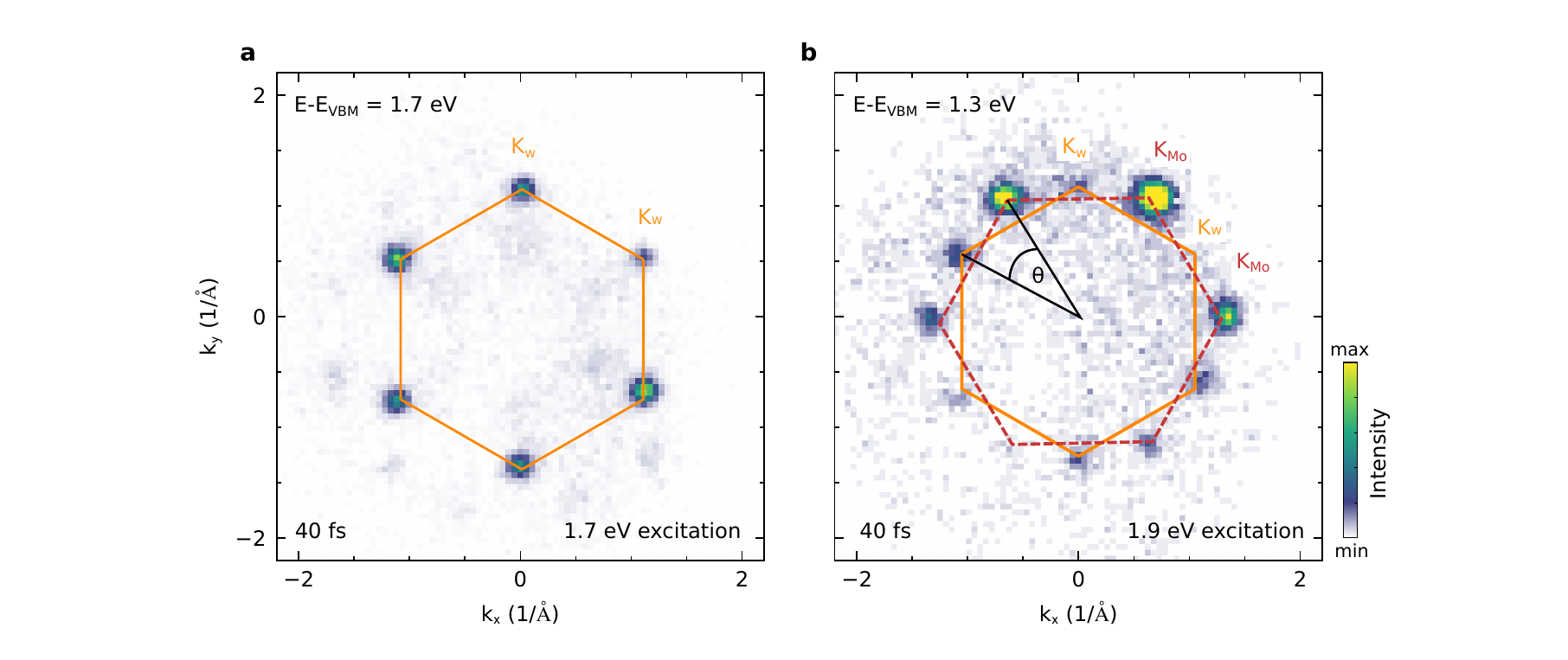}
    \caption{\textbf{Determination of the twist angle of the WSe$_2$/MoS$_2$ heterostructure.}
    Momentum fingerprints of the WSe$_2$ A$_{\rm W}$-exciton \textbf{a} and the MoS$_2$ A$_{\rm Mo}$-exciton \textbf{b} after photoexcitation with 1.7~eV and 1.9~eV, respectively (pump-probe delay of 40~fs). The Brillouin zones of WSe$_2$ (orange) and MoS$_2$ (dark red, dashed) are overlaid onto the data as hexagons. The twist angle $\theta$ can be directly determined from the misalignment of the WSe$_2$ and MoS$_2$ K valleys (labelled as K$_{\rm W}$ and K$_{\rm Mo}$). } 
    \label{twistangle}
\end{figure}

The WSe$_2$/MoS$_2$ heterostructure is fabricated using mechanical exfoliation and dry transfer methods, as detailed in ref.~\cite{Schmitt22nat}. The heterostructure is stamped on $\approx 30$~nm hBN~\cite{Taniguchi07jcg} on top of a p$^+$-doped silicon wafer with a native oxide layer. Real-space images of the sample are shown in Fig.~1b,c of the main text. Before the photoemission experiments, the sample is annealed under ultra-high vacuum conditions to a temperature of 670 K for 1 hour. The twist-angle of the heterostructure is determined to 28.8$\pm$0.8° by direct comparison of the momentum misalignment of the excitonic photoemission signals from WSe$_2$ and MoS$_2$ A1s excitons (Extended Data Fig.~\ref{twistangle}, cf. analysis in ref.~\cite{Schmitt22nat}).

\clearpage

\section{ILX formation mechanism: Two-step process via layer hybridized excitons}

\begin{figure}[htb!]
    \centering
    \includegraphics[width=\linewidth]{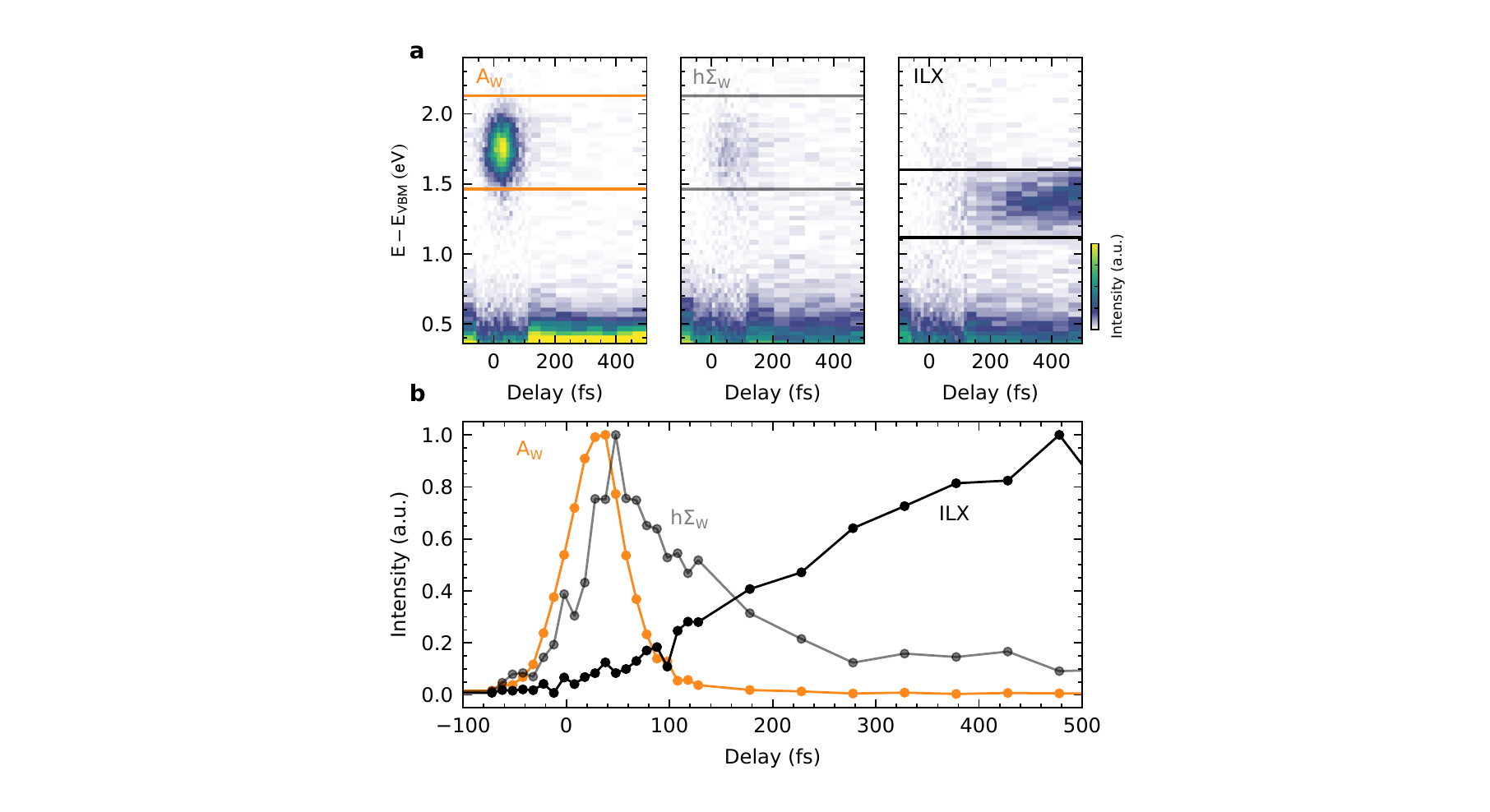}
    \caption{\textbf{Characterization of the ILX formation mechanism and analysis of the pump-probe-delay-dependent photoemission yield from excitons.}
    \textbf{a} Pump-probe delay evolution of energy-distribution curves (EDCs) extracted on the momenta of the optically-excited A$_{\rm W}$-exciton (left panel), the hybrid h$\Sigma$-exciton (middle panel) and the ILX (right panel).
    \textbf{b} By integrating the spectral weight in the energy-windows indicated by the colored rectangles in \textbf{a}, the delay-dependent photoemission signal can be directly compared. A temporal hierarchy of photoemission yield  from A$_{\rm W}$-excitons (orange), h$\Sigma$-excitons (grey) and ILX (black) is observed. Note, that panel \textbf{a} shows unnormalized EDCs and \textbf{b} normalized delay-dependent photoemission signals.} 
    \label{momentum_resolved_dynamics}
\end{figure}

For the 28.8°$\pm$ 0.8° twisted WSe$_2$/MoS$_2$ heterostructure, the formation mechanism of the ILX is a priori not clear. Because of the large twist angle, an important question is if the ILX formation is still mediated by interlayer hybridized h$\Sigma$-excitons, as recently shown for a 9.8$\pm$0.8° twisted heterostructure in ref.~\cite{Schmitt22nat}. In order to address this question, we evaluate the pump-probe delay evolution of all observed excitonic photoemission signals, including the A$_{\rm W}$-exciton (orange), the h$\Sigma$-exciton (grey) and the ILX (black)  (Extended Data Fig.\ref{momentum_resolved_dynamics}). At about 30~fs, we find a peak in the photoemission intensity from optically excited A$_{\rm W}$ excitons, which is followed by a fast decrease of the signal. During the rise and decrease of the A$_{\rm W}$-exciton occupation, spectral yield of the h$\Sigma$-exciton builds up (grey symbols). After reaching its maximum at around 50 fs, the signal decreases slowly, and, concomitant, the photoemission yield from the ILX increases and saturates on the 500~fs timescale (black symbols). This hierarchy of timescales was also found in refs.~\cite{Schmitt22nat,Bange23arxiv2} and is consistent with the two-step ILX formation process via h$\Sigma$-excitons, i.e., via the cascade A$_{\rm W}\rightarrow$h$\Sigma\rightarrow$ILX.



\clearpage


\section{AFM measurements on the WSe$_2$/MoS$_2$ heterostructure}

\begin{figure}[htb!]
    \centering
    \includegraphics[width=\linewidth]{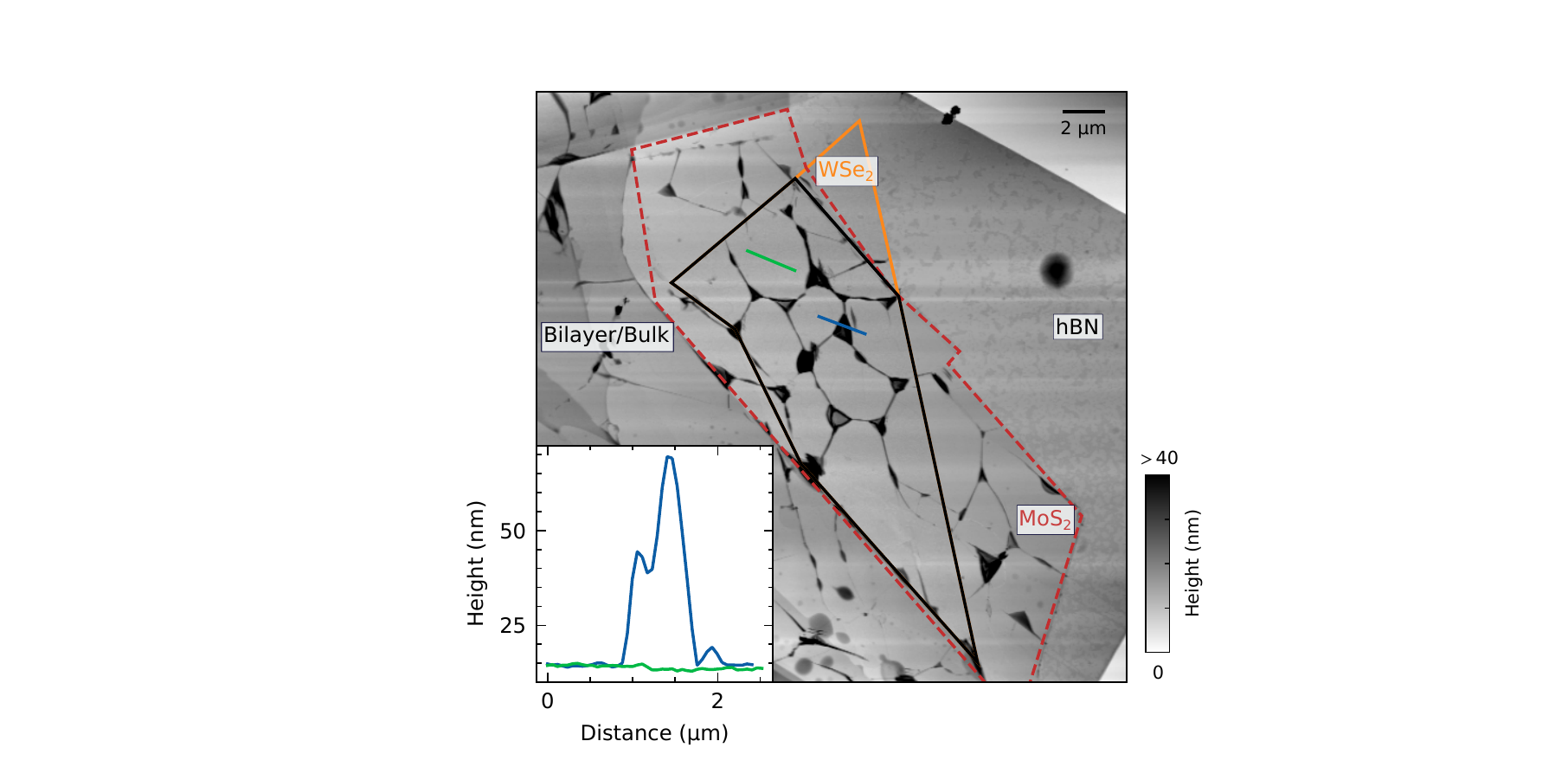}
    \caption{\textbf{AFM image of the moiré heterostructure}. The WSe$_2$ and MoS$_2$ monolayers and the WSe$_2$/MoS$_2$  heterostructure  are indicated by orange, dark-red (dashed) and black polygons, respectively. In addition, the bulk and the hBN regions are labelled. The inset shows line profiles taken across a blister (blue line) and across a smooth sample area (green line) of the heterostructure.}
    \label{afmimage}
\end{figure}

AFM measurements are performed to characterize the real-space homogeneity of the WSe$_2$/MoS$_2$ heterostructure (Extended Data Fig.~\ref{afmimage}). The AFM-image is measured under ambient conditions after the sample has experienced the annealing procedure for the photoemission measurement in ultra-high vacuum conditions. We can clearly resolve the WSe$_2$ (orange polygon) and MoS$_2$ (dark red polygon) monolayer regions and their area of overlap. In addition, next to the monolayer flakes, trilayer and bulk regions are found that can be distinguished from the monolayer regions by larger topographic heights. We observe that the heterobilayer region exhibits irregularities with heights up to 100~nm (inset, blue line profile). These areas are identified, for the most part, as residual gas and hydrocarbons being trapped at the interface between the WSe$_2$ and MoS$_2$ layers or the MoS$_2$ and hBN layers. Importantly, in the ultrafast dark-field momentum microscopy experiments, we exclude these blister areas from the analysis  (cf. AFM overlays in Fig.~2 and 4 of the main text) by removing all regions from the dark-field data with AFM heights larger than 20 nm. Apart from these local areas, the heterobilayer is comparably flat with height differences in the order of 1-2~nm (inset, green line profile). 

\clearpage

\section{Experimental setup: ultrafast dark-field momentum microscopy}

The dark-field momentum microscopy experiments are performed with a time-of-flight momentum microscope~\cite{medjanik_direct_2017} (Surface Concept, ToF-MM) that is operated with a table-top high-repetition rate high-harmonic generation beamline. The beamline is based on a 300~W AFS Fiber laser system and the pump photon energy is tuneable to the bright exciton resonance with an optical parametric amplifier (Orpheus-F/HP from Light Conversion)~\cite{Keunecke20timeresolved,Keunecke20prb,Duvel22nanolett}. 

All experiments are performed at room temperature with 26.5~eV probe photons ($p$-polarized, 21$\pm$5~fs) and 1.7~eV pump photons ($s$-polarized, 50$\pm$5~fs). 
The applied pump fluence is 340$\pm$20~µJ/cm$^{-2}$ and, following the analysis of ref. \cite{Schmitt22nat}, the absorbed pump fluence is 1.7$\pm$0.3~µJ/cm$^{-2}$ corresponding to an exciton density of (6.5$\pm$1.0) $\times$ $10^{12}$~cm$^{-2}$.

\subsection{Data handling and image correction}

In the dark-field momentum microscopy experiment, we monitor the real-space-resolved exciton dynamics of the A$_{\rm W}$-exciton or the ILX by varying the delay between the pump- and the probe laser pulse. As the dark-field aperture blocks a significant amount of photoelectrons, longer integration times are necessary. In the case of the measurement of the ILX dynamics, this implies that we perform overall 23 measurement cycles, whereas, in each cycle, the pump-probe delay is varied between 0~fs and 500~fs in 50~fs steps (10~min integration time per delay step). Overall, this adds up to a measurement duration of 53~h. The overall integration time for the measurement of the A$_{\rm W}$-exciton dynamics is 21~h.


\subsubsection{Correction of rigid energy-shifts}

As typical for measurements on exfoliated TMDs stamped onto Si substrates with a native oxide layer, a pump-probe delay-dependent rigid shift of the full energy spectrum is observed. This rigid energy shift can be attributed to space-charge or surface photovoltage effects~\cite{Schonhense21rsi}. We correct this energy shift for each measurement cycle, as we have detailed in refs.~\cite{Schmitt22nat,Bange23arxiv}.


\subsubsection{Correction of rigid real-space shifts}

In addition, during the long integration times, rigid shifts of the real-space resolved data can occur due to slight movements of the sample in the momentum microscope. In order to correct for such shifts, we calculate a cross-correlation of each real-space resolved photoelectron image with a reference image. Subsequently, we correct for the rigid real-space shifts before merging the individual data sets.

\subsubsection{Correction of image distortions}

Momentum microscopy data can be affected by different types of lens errors that can lead to various image distortions such as stretching, shearing or barrel distortions \cite{xian2020open,Karni22nat}. Moreover, by inserting a contrast-aperture into the Fourier plane of the microscope, off-axis photoelectrons are filtered that are then projected onto the photoelectron detector with an energy-dependent lateral shift. While these distortions can be partly corrected in experiment with magnetic deflectors, further post-processing steps of the measured data are necessary~\cite{Keunecke20timeresolved,xian2020open}. In conventional momentum-resolved measurements, such corrections are done by a combination of (1) symmetrization of the photoemission signatures (e.g., the known dimension of the Brillouin zone)~\cite{xian2019symmetry} and/or (2) the symmetrization with a grid which can be placed in the Fourier plane of the objective lens~\cite{Maklar20rsi, Karni22nat}. 

For the dark-field momentum microscopy experiments, a correction via symmetry-arguments (1) is not possible. A correction by placing a grid in the real-space plane (2) would be possible, but would only correct the lens errors of the projective lens, but not of the objective lens. For our correction procedure, we first (i) correct for the energy-dependent lateral shift and second (ii) use an AFM-image of the heterobilayer sample as a reference image to align the real-space photoelectron images. 

First (i), for the correction of the energy-dependent lateral shift, we quantitatively analyze this shift based on a cross-correlation between images at different photoelectron energies. Note that we observe a linear lateral shift for increasing energy, which is then corrected for every image.

\begin{figure}[htb!]
    \centering
    \includegraphics[width=\linewidth]{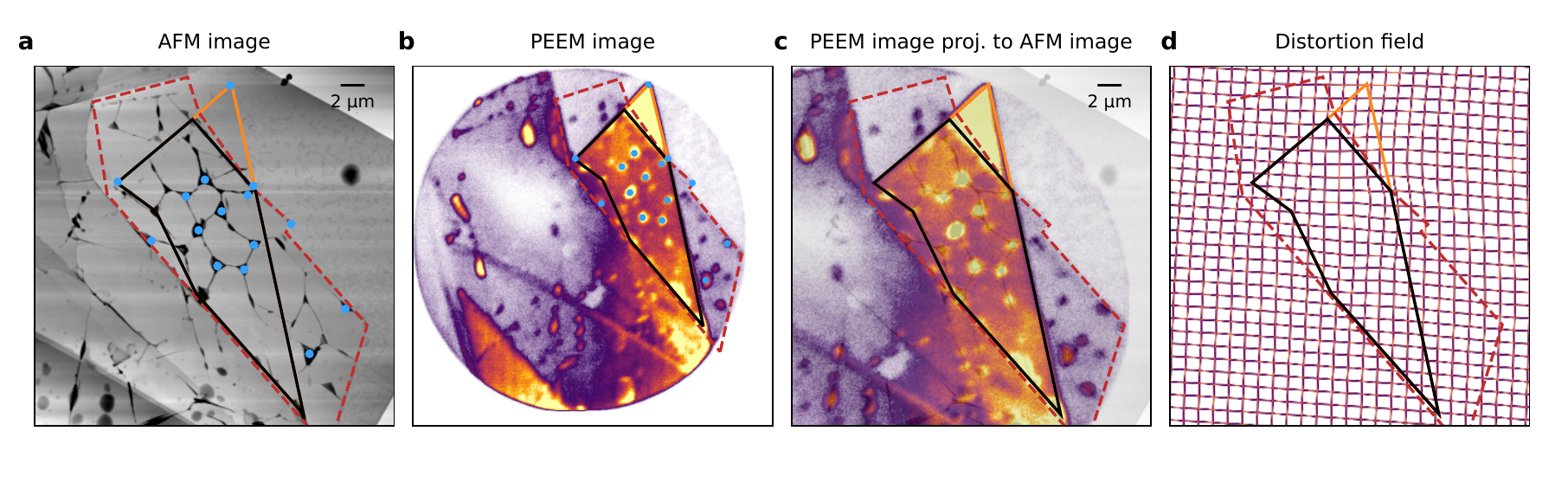}
    \caption{\textbf{Image correction in dark-field momentum microscopy.}
    \textbf{a} The WSe$_2$ (orange) and MoS$_2$ (dark red, dashed) monolayers and the WSe$_2$/MoS$_2$-heterostructure (black) and blisters (blue points) can be identified in the 30~µm$\times$30~µm AFM image. 
    \textbf{b} Exemplary real-space image of the full field-of-view measured with the momentum microscope at an energy of E-E$_{\rm VBM,avg}$ = -0.5~eV. The dark-field aperture is positioned at the in-plane momenta of the ILX. The boundaries of the WSe$_2$ and MoS$_2$ monolayers as well as the blister positions are indicated.
    \textbf{c} Based on an affine transformation, the AFM and photoemission images are aligned.
    \textbf{d} Distortion field of the applied transformation.}
    \label{distortionfield}
\end{figure}

The distortion correction (ii) is performed by selecting distinctive points of the heterostructure in the AFM and the photoelectron image (e.g., blisters and edges, Extended Data Fig.\ref{distortionfield}). Thereafter, we apply an affine transformation that projects the real-space points of the photoelectron image at a certain energy onto respective coordinates extracted from the AFM image; the respective distortion field is visualized in Extended Data Fig.~\ref{distortionfield}. It is important to note that the energy-dependent shift is corrected before the projection is performed. In that way, the same projection can be applied for each energy in the data set.


\clearpage


\section{Calibration of the energy-axis and pump-probe-delay-axis in dark-field momentum microscopy}

In addition to the dark-field momentum microscopy experiment, we have performed a typical momentum-resolved photoemission experiment on the WSe$_2$/MoS$_2$ heterostructure (cf. Extended Data Fig.\ref{momentum_resolved_dynamics}). We use this spatially averaged trARPES experiment to calibrate the energy- and the pump-probe-delay-axis of the dark-field momentum microscopy experiment.

\subsection{Energy axis}
\label{sec:energyaxis}

In most ARPES-related studies on TMDs, the topmost valence-band is set as reference point for the energy axis. For this, the exact position of the VBM must be accessible in experiment. However, this is not straightforwardly realized in dark-field momentum microscopy: The dark-field aperture has an effective diameter of 0.4~\AA$^{-1}$  (blue circle in Extended Data Fig.\ref{energy-time-cali}a). As such, it integrates over a broad momentum region of the hole-like WSe$_2$ valence band, and, in consequence, the maximum of the photoemission intensity shifts to smaller energies. This is illustrated in Extended Data Fig.\ref{energy-time-cali}a and b where, from typical momentum-resolved measurements, EDCs are evaluated for different apertures sizes, ranging from the original aperture size of dark-field experiment (dark-field apert. size x 1.00 in Extended Data Fig.\ref{energy-time-cali}a and b) to roughly 5 times smaller aperture size (dark-field apert. size x 0.21 in Extended Data Fig.\ref{energy-time-cali}a and b). From a gaussian peak-fitting procedure, we find an energetic offset of the VBM of 0.23 eV between the respective aperture sizes. In order to provide a meaningful and comparable energy scale in the dark-field momentum microscopy experiment, we shift all data sets with this constant offset. This allows the experiment to provide correct exciton energies that are comparable to other experiments.   

Note that the lateral shifts of the energies of the top of the WSe$_2$ VB as discussed in the main text (Fig.~4a,c) lead, in principle, to a spatially varying energy scale of the plotted data. In consequence, in Fig.~4a,b, we plot the data with respect to the spatially averaged VBM, i.e., with respect to E$_{\rm VBM,avg}$. 




\subsection{Alignment of the delay axis for the ILX and the A$_{\rm W}$-exciton dark-field momentum microscopy experiment}

\begin{figure}[htb!]
    \centering
    \includegraphics[width=1\linewidth]{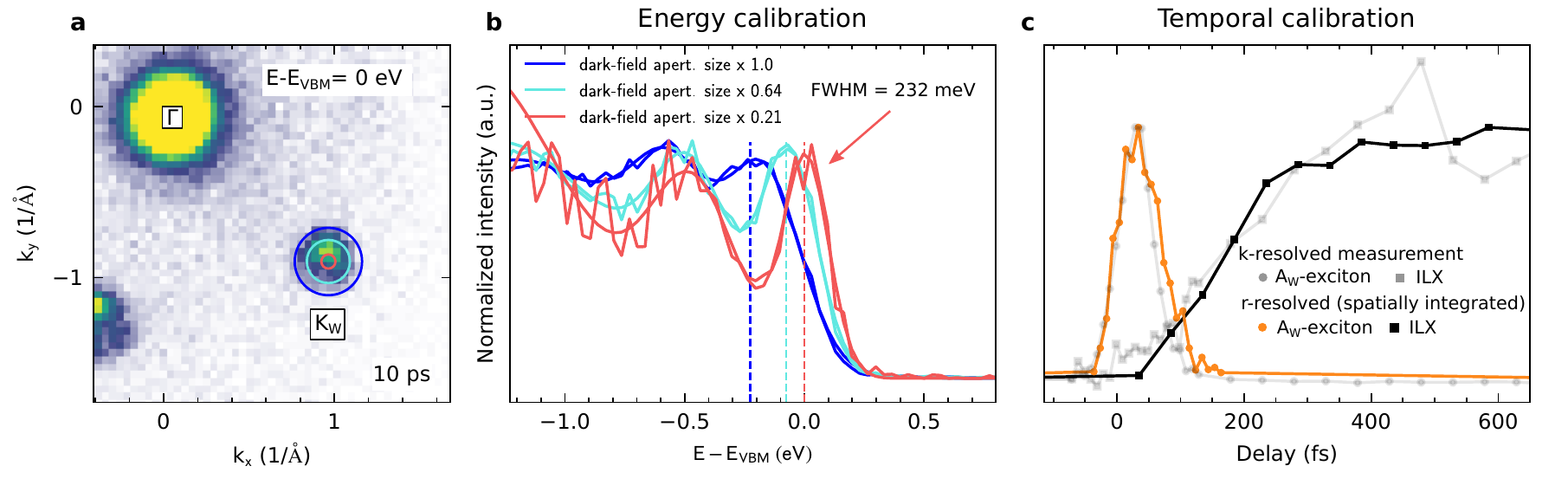}
    \caption{\textbf{Calibration of the energy- and the pump-probe-delay axis in the dark-field momentum microscopy experiment.}
    \textbf{a} In-plane momentum-resolved photoemission data taken at a pump-probe delay of 10 ps and the energy centered on the WSe$_2$ VBM. The blue, cyan and red circle indicate different aperture sizes used to create the EDCs in \textbf{b}. The size of the dark-field aperture corresponds to the blue circle.
    \textbf{b} EDCs obtained by integrating the photoemission yield in the respective momentum-regions of interest indicated in \textbf{a}. In all EDCs, the energetically highest photoemission peak is attributed to the WSe$_2$ VBM. With increasing diameter of the aperture in momentum space, the peak maxima shift to smaller energies. A quantitative analysis shows an energetic offset of 0.23~eV between the 1$\times$ aperture (i.e., the size of the dark-field aperture) and 0.21$\times$ aperture. The peak position of the 0.21$\times$ aperture is set as reference point and the spatial-resolved data is calibrated with respect to the extracted offset-value. 
    \textbf{c} Pump-probe delay-dependent analysis of the photoemission intensity of A$_{\rm W}$-excitons and ILXs as extracted in the dark-field experiment (orange and black) and the spatially averaged trARPES experiment (grey).}
    \label{energy-time-cali}
\end{figure}

As the real-space resolved dynamics of the ILX and the A$_{\rm W}$-exciton are obtained in separate experiments with long integration times, the pump-probe delay axis of both measurements has to be matched. We do this by directly comparing the dark-field experiments with the spatially averaged trARPES experiment. Therefore, in the spatially averaged trARPES experiment, we evaluate the pump-probe delay-dependent photoemission intensity from the bright A$_{\rm W}$-exciton and the ILX as shown in Extended Data \ref{momentum_resolved_dynamics} (Extended Data Fig.\ref{energy-time-cali}c, grey data points). Importantly, in this analysis, the dynamics of the A$_{\rm W}$-exciton and the ILX are measured simultaneously, and, hence, the experiment provides the real delayed onset of the ILX with respect to the A$_{\rm W}$ exciton. In order to align the delay axis of the ILX and the A$_{\rm W}$-exciton photoemission yield in the dark-field experiment, we spatially integrate photoelectron counts obtained from the ILX (on the heterobilayer region, black squares) and the A$_{\rm W}$-exciton (on the heterobilayer region, orange points) and align those with the dynamics in the momentum-resolved trARPES experiment (grey points and squares)  in Extended Data Fig.\ref{energy-time-cali}c, respectively. Note that this alignment of the pump-probe-delay axis does not impact the quantitative evaluation of the ILX formation time, because here only the formation time is extracted. 

\clearpage


\section{Additional spatio-temporal snapshots and comparison to the exciton dynamics in the momentum-resolved experiment.}

\begin{figure}[htb!]
    \centering
    \includegraphics[width=\linewidth]{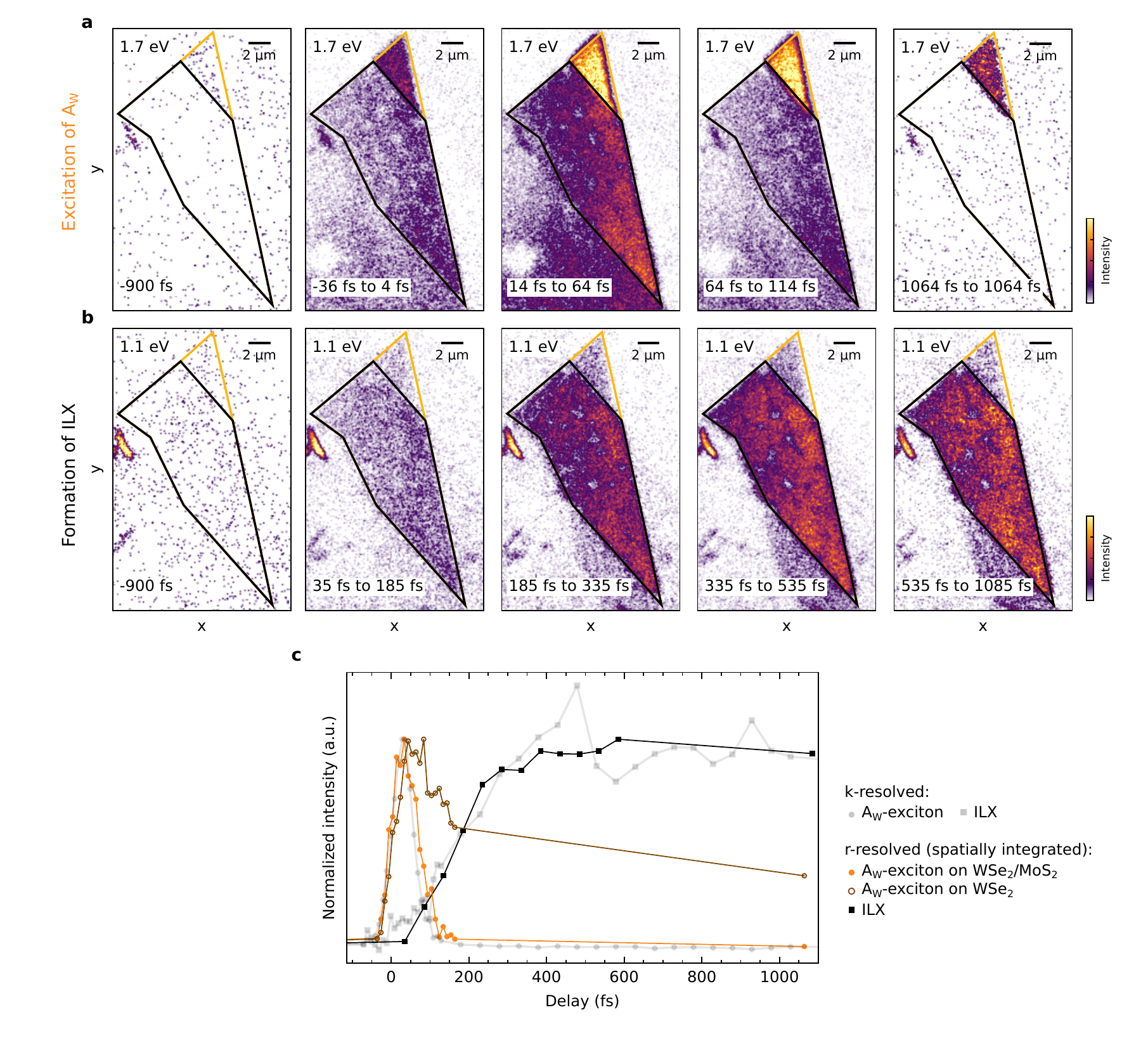}
    \caption{\textbf{a, b} Real-space snapshots of the formation and relaxation dynamics of bright A$_{\rm W}$-excitons (\textbf{a}) and ILXs (\textbf{b}).
    \textbf{c} Pump-probe delay-dependent analysis of the photoemission yield of the A$_{\rm W}$-exciton and the ILX in the WSe$_2$/MoS$_2$ heterobilayer (orange and black) and the WSe$_2$ monolayer (brown). The data points are obtained in the dark-field momentum microscopy experiment by integrating photoelectron counts inside the heterobilayer and monolayer regions. The grey data points are generated from a spatially-averaged trARPES experiment with an aperture positioned in the real-space plane of the microscope (cf. blue circle in Fig.~1b).}
    \label{compa-real-momentum-dynamics}
\end{figure}

As an extension to Fig.~2, Extended Data Fig.\ref{compa-real-momentum-dynamics} shows additional spatio-temporal snapshots of the A$_{\rm W}$-excitons and the ILX. By spatially integrating over the WSe$_2$ monolayer region and the WSe$_2$/MoS$_2$ heterobilayer region of the A$_{\rm W}$ signal (orange and black region-of-interest in Extended Data Fig.\ref{compa-real-momentum-dynamics}a), we get direct access to the A$_{\rm W}$-exciton and ILX dynamics on monolayer WSe$_2$ and heterobilayer WSe$_2$/MoS$_2$ (Extended Data Fig.\ref{compa-real-momentum-dynamics}c). 


\clearpage

\section{Quantitative analysis of the ILX formation time and the exciton energies}

Figures 2 and 4 of the main text quantitatively analyze the ILX formation time and the excitonic energy landscape.

\subsection{Quantitative analysis of the ILX formation time}
\label{sec:errorfunctionfit}

The rise of photoemission yield of ILX is fitted with error functions of the form
\begin{equation}
    I/I_{\rm max}=0.5\cdot\left({\rm erf}\left(\frac{t-t^*}{\sqrt2 w}\right)+1\right),
    \label{eq:errorfunction}
\end{equation}
where $I/I_{\rm max}$ describes the photoelectron intensity normalized to the maximum intensity $I_{\rm max}$, $t^*$ the onset and $w$ the width of the errorfunction. We fit the temporal evolution in the regions-of-interest with this error function and evaluate its full-width at half-maximum value ($w \times 2\sqrt{2\ln{2}}$) to quantify the ILX formation time.

\subsection{Quantitative analysis of the exciton energies}
\label{sec:determinationEXCenergy}

Recent reports have shown that the exciton energy E$_{\rm exc}^i$ of bright and dark excitons can be extracted from trARPES experiments by considering the conservation of energy when the Coulomb correlation between the exciton's electron and hole is broken in the photoemission process~\cite{Weinelt04prl,Bange23arxiv,Bennecke23arxiv,Christiansen19prb,Rustagi18prb}. This statement is well-described by the relationship
\begin{equation}
    E_{\rm elec}=E_{\rm hole}+E^i_{\rm exc}+\hbar\omega,
    \label{eq:energyconser}
\end{equation}
which shows that the photoemission energy of the single particle electron E$_{\rm elec}$ is detected one exciton energy E$_{\rm exc}^i$ above the energy of the single particle hole E$_{\rm hole}$  that remains in the sample ($\hbar\omega$: photon energy). When probing photoelectrons from A$_{\rm W}$-excitons and the ILXs, the exciton's hole remains in the VBM of WSe$_2$. Hence, we can extract the spatio-spectral dependence of A$_{\rm W}$-excitons and ILXs by fitting the photoemission signals from A$_{\rm W}$-excitons, ILX, and the WSe$_2$ valence bands with Gaussian functions. Subsequently, we calculate the energy difference between the excitonic photoemission signal and the WSe$_2$ VBM maxima. This is indicated, exemplarily, by the double-headed arrow in the EDCs in Fig.~4a. Importantly, as discussed in section~\ref{sec:energyaxis}, we correct the extracted energies by a constant energy offset of 0.23~eV because of the momentum-broadening of the EDCs induced by the finite size dark-field aperture.

\clearpage


\section{Real-space resolved heatmaps of the formation time, the energy landscape and the respective absolute errors}

\begin{figure}[htb!]
    \centering
    \includegraphics[width=\linewidth]{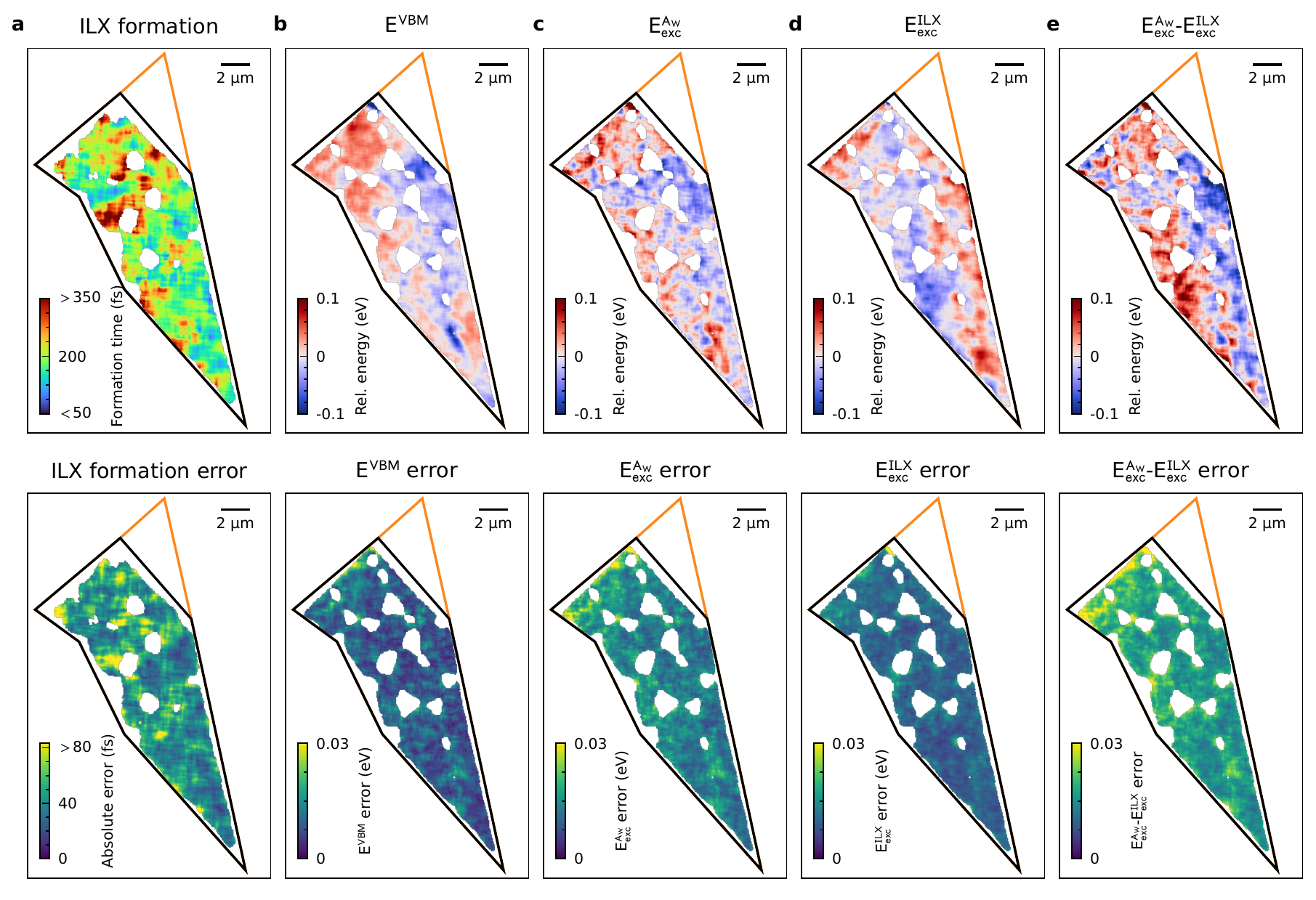}
    \caption{\textbf{Real-space resolved heatmaps of the formation time, the energy landscape and the respective errors.}
    \textbf{a} ILX formation map and respective absolute errors. \textbf{b-e} Spatio-spectral heatmaps (top row) and respective absolute error maps (bottom row)}
    \label{errormaps}
\end{figure}

In Figs.~2 and 4 of the main text, heat-maps that illustrate the formation time of ILX and the electronic and excitonic energy landscape are shown. In Extended Data Fig.\ref{errormaps}, the heat-maps are reproduced next to maps that show the spatially resolved absolute errors. For the formation time of ILX, this error is obtained from the fitting of the pump-probe delay-dependent photoemission intensity with equation~\ref{eq:errorfunction}. For the energy landscape, the errors are obtained from Gaussian fits to the photoemission feature after the analysis based on equation~\ref{eq:energyconser}.

\clearpage


\section{Spatial resolution of the ultrafast dark-field momentum microscopy experiment} 

\begin{figure}[htb!]
    \centering
    \includegraphics[width=\linewidth]{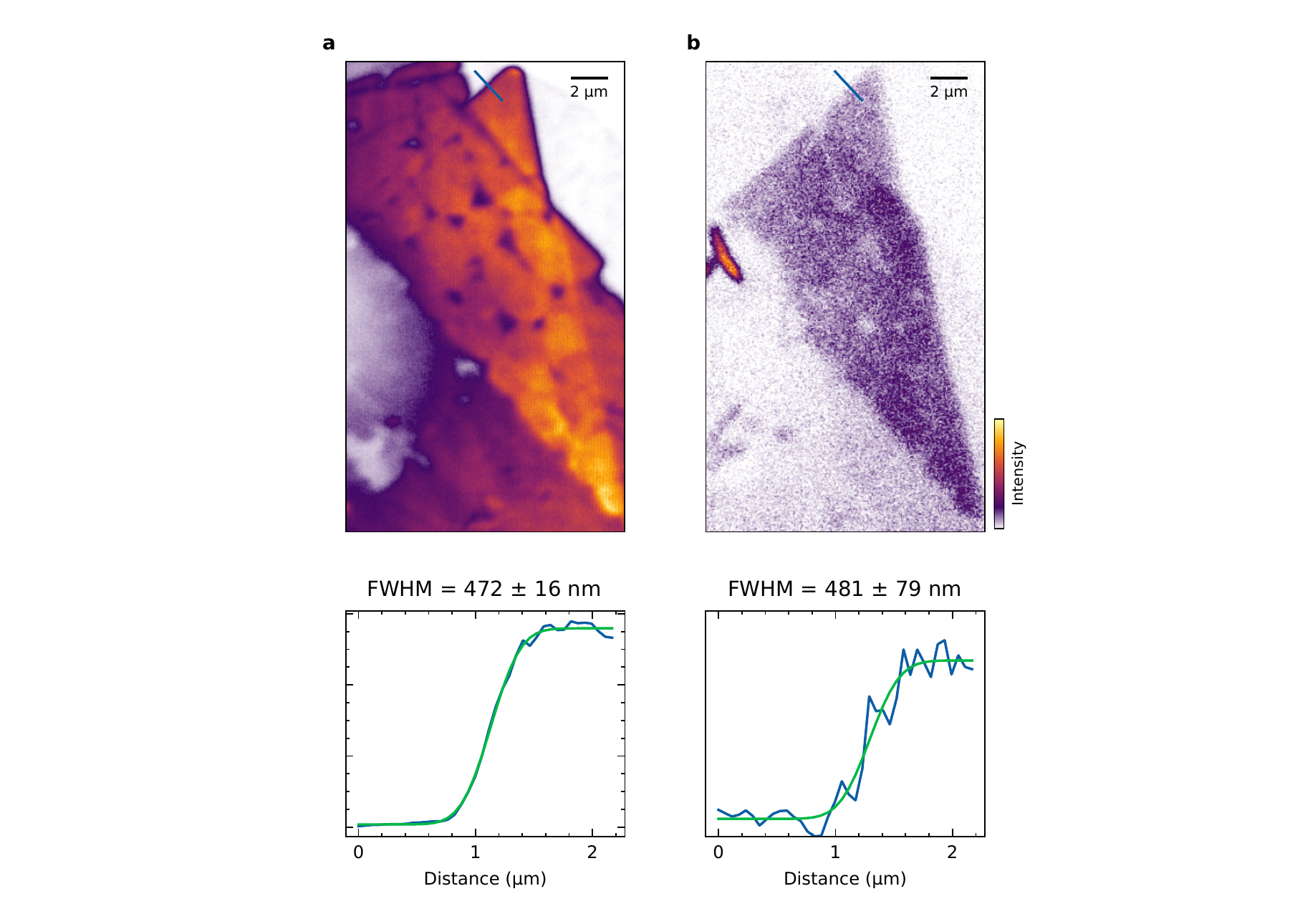}
    \caption{\textbf{Benchmarking of the real-space resolution of the dark-field momentum microscopy experiment.}
    \textbf{a} In the momentum microscopy experiment with the dark-field aperture placed on the in-plane momenta of the ILX at an energy range from -1.69~eV to -1.49 eV (occupied valence bands) and a photon energy of 26.5~eV, the spatial resolution is determined to 472$\pm$16~nm. 
    \textbf{b} If the same analysis is done on the excitonic photoemission yield, i.e., integrated over all energies above 0.6~eV, a spatial resolution of 481$\pm$79~nm is extracted. All green lines in the lower panels are error function fits, from which we extract the FWHM has the spatial resolution.}
    \label{resolution-real-space}
\end{figure}

We quantify the real-space resolution of the dark-field momentum microscopy experiment to 472$\pm$16~nm (for the used settings). This value is obtained by fitting the change in the photoemission intensity across the lateral interface from bulk hBN to monolayer WSe$_2$ with an error function (Extended Data Fig.\ref{resolution-real-space}a, FWHM). Notably, this spatial-resolution is obtained when the experiment is performed with 26.5~eV photons, the dark-field aperture is positioned on the ILX photoemission signal and photoelectron counts from the occupied valence bands are evaluated in an energy window from 1.69~eV to 1.49~eV. Alternatively, if we directly evaluate photoemission yield from excitons along the same direction on the sample, we find a comparable spatial resolution of 481$\pm$79~nm (FWHM) for our settings at a reduced signal-to-noise ratio (Extended Data Fig.\ref{resolution-real-space}b). We note that the momentum microscopy instrument itself is in principle capable to achieve a spatial resolution well below 100~nm.


\clearpage






\clearpage

\section{Microscopic modelling}
\label{sec:theory}

We introduce the main theoretical steps needed in order to access the hybrid exciton dynamics in TMD bilayers. The Hamiltonian of the system can be written as 
\begin{equation}\label{eq:4}
    H = H_0 + H_T = 
    \sum_{\mu, {\bf Q}}{ E^{\mu }_{{\bf Q}} X^{\mu\dagger}_{{\bf Q}} X^{\mu}_{{\bf Q}} }
    + \sum_{\substack{\mu,\nu,{\bf Q}}}{ \Tb_{\mu \nu}  {X^{\mu\dagger }_{{\bf Q}}}X^{\nu}_{{\bf Q}} }
\end{equation}
whith superindex $\mu = (n^\mu, \zeta^\mu_e,\zeta^\mu_h,L^\mu_e,L^\mu_h)$ describing the exciton degrees of freedom. Furthermore, $E^\mu_{\bf Q} = E^c_{\zeta^\mu L^\mu_e}-E^v_{\zeta^\mu L^\mu_h} + E^\mu_{bind} +  E^\mu_{ {\bf Q},kin}$ are the excitonic energies, where $E^\mu_{bind}$ are obtained after solving a bilayer Wannier equation \cite{ovesen2019interlayer, brem2020hybridized}. Moreover, $E^{c/v}_{\zeta^\mu L^\mu_e}$ is the conduction and valence band energy and $E^\mu_{{\bf Q},kin} = \hbar^2 {\bf Q}^2/ (2 M^\mu)$ is the kinetic energy of the exciton with mass $M^\mu =(m^\mu_e+m^\mu_h)$. We address the excitonic tunneling between the TMD monolayers by starting from the electronic tunnelling matrix elements
\begin{equation}\label{eq:5}
   \Tb_{\mu \nu} = (\delta_{L^{\mu}_h L^{\nu}_h} (1- \delta_{L^{\mu}_e L^{\nu}_e})  \delta_{\zeta^\mu \zeta^\nu} T^c_{\mu_e,\nu_e} - \delta_{L^{\mu}_e L^{\nu}_e} (1- \delta_{L^{\mu}_h L^{\nu}_h}) \delta_{\zeta^\mu \zeta^\nu} T^v_{\mu_h,\nu_h})  \sum_{\bf k}{ \psi^{\mu*}({\bf k} )\psi^{\nu}({\bf k}) },
\end{equation}
where $\psi^{\mu}$ is the excitonic wave function of the state $\mu$ defined over the relative momentum between electron and hole. Here, $T^\lambda_{i j} = \bra{\lambda i {\bf p} } H \ket{\lambda j {\bf p}} (1-\delta_{L_i L_j}) \delta_{\zeta_i \zeta_j}$ denotes the electronic tunneling elements obtained by averaging DFT values of MoSe$_2$-WSe$_2$ and MoS$_2$-WS$_2$ heterostructures in \cite{hagel2021exciton}. 
Diagonalizing Eq. \ref{eq:4} leads to a new set of hybrid excitonic energies $\Ea^\eta_{\bf Q}$ that are obtained by solving the hybrid eigenvalue equation \cite{brem2020hybridized, hagel2021exciton},
\begin{equation}\label{eq:6}
    E^\mu_{\bf Q} c^\eta_\mu({\bf Q}) + \sum_{\nu}{ \Tb^{}_{\mu \nu} c^\eta_\nu({\bf Q}) } = \Ea^\eta_{\bf Q} c^\eta_\mu({\bf Q}).
\end{equation}
The diagonilized hybrid exciton Hamiltonian reads \cite{Meneghini22naturalsciences,Schmitt22nat}
\begin{equation}\label{eq:7}
     H= \sum_{\eta}{\Ea^\eta_{\bf Q}  Y^{\eta \dagger}_{\bf Q}  Y^\eta_{\bf Q} } 
\end{equation} 
with the hybrid exciton annihilation/creation operators $Y^{\eta(\dagger)}_{\bf Q} = \sum_{ \mu }{ c^\eta_\mu({\bf Q}) X^{\mu(\dagger)}_{\bf Q}}$.
Using \ref{eq:7} we have access to the hybrid exciton energy landscape for the investigated  WSe$_2$-MoS$_2$ heterostructure. 

When treating hybrid states consisting of two main contributions (intra- and interlayer state of one species), and considering that $\abs{c^\eta_\mu({\bf Q})}^2$ is related to the percentage of intra- or interlayer character inside the hybrid state, it is useful to define a new quantity called degree of hybridization (DoH)
\begin{equation}\label{eq:doh}
     DoH(Q) = 1- \abs{\abs{c^\eta_{intra}({\bf Q})}^2 - \abs{c^\eta_{inter}({\bf Q})}^2} 
\end{equation} 
The DoH has maximum at 1 when the new state is maximally hybridized (50-50\% mixture) between the two starting states, and minimum at 0 in the extreme case of purely intra/interlayer state.

The hybrid exciton-phonon scattering plays a crucial role at the low excitation regime \cite{brem2018exciton,Meneghini22naturalsciences}. The  corresponding Hamiltonian can be written as \cite{brem2020hybridized}
\begin{equation}\label{eq:9}
    H_{Y-ph} = \sum_{j,{\bf Q},{\bf q}, \eta,\xi}{ \tilde{\Da}^{\xi\eta}_{j,{\bf q},{\bf Q}} Y^{\xi\dagger }_{{\bf Q + q}}Y^{\eta}_{{\bf Q}} b_{j,{\bf q}} } + h.c.
\end{equation}
with the hybrid exciton-phonon coupling $\tilde{\Da}^{\xi\eta}_{j,{\bf q},{\bf Q}}$. The electron-phonon matrix elements, single-particle energies and effective masses are taken from DFT calculations \cite{PhysRevB.90.045422}.
The excitation of the system through a laser pulse is described semi-classically via the minimal-coupling Hamiltonian that can be written as \cite{brem2020hybridized}
\begin{equation}
    H_{Y-l} = \sum_{\sigma,{\bf Q},\eta}{ {\bf A} \cdot \tilde{\mathcal{M}}^\eta_{\sigma {\bf Q}} Y^\eta_{\bf Q_{\parallel} }} + h.c.\\
\end{equation}
with hybrid exciton-light coupling $\tilde{\mathcal{M}}^\eta_{\sigma {\bf Q}}$.
Details on the transformation and the definition of the hybrid interaction matrix elements and couplings are given in Ref. \cite{brem2020hybridized, hagel2021exciton}.

Direct access to the dynamics of the system is obtained by solving the Heisenberg equation of motion for the hybrid occupation $N^\eta = \langle Y^{\eta\dagger }Y^{\eta}  \rangle$, including $H = H_Y + H_{Y-ph}+ H_{Y-l}$, and truncating the Martin-Schwinger hierarchy using a second order Born-Markov approximation \cite{kira2006many,haug2009quantum,malic2013graphene}, separating coherent $P^{\eta}_{\bf Q} =\langle Y^{\eta \dagger}_{\bf Q}\rangle$ and incoherent hybrid populations $\delta N^{\eta }_{\bf Q} = \langle Y^{\eta\dagger }_{{\bf Q}}Y^{\eta}_{{\bf Q}}\rangle-\langle Y^{\eta \dagger}_{\bf Q}\rangle \langle Y^{\eta}_{\bf Q} \rangle = N^{\eta }_{\bf Q} - |P^{\eta}_{\bf Q}|^2$. This leads to the following semiconductor Bloch equations
\begin{align}\label{eq:10}
\begin{split}
    i\hbar \partial_t P^\eta_0 &= -(\Ea^\eta_0 + i \Gamma^\eta_0)P^\eta_0 -  \tilde{\mathcal{M}}^\eta_0 \cdot {\bf A}(t)\\[6pt]
    \delta \dot{N}^\eta_{\bf Q} &= \sum_{\xi}{ W^{\xi\eta}_{{\bf 0 Q}}  \abs{P^{\eta}_0}^2 } + \sum_{\xi, {\bf Q'}}{ \left( W^{\xi\eta}_{{\bf Q' Q}} \delta N^\xi_{\bf Q'} - W^{\eta\xi}_{{\bf Q Q'}} \delta N^\eta_{\bf Q} \right) }
\end{split}
\end{align}
 with $W^{\eta\xi}_{{\bf Q Q'}} = \frac{2\pi}{\hbar} \sum_{j,\pm}|\Da^{\eta\xi}_{j,{{\bf Q'-Q}}} |^2 \left( \frac{1}{2} \pm \frac{1}{2} + n^{ph}_{j,{\bf Q'-Q}} \right) \delta \left( \Ea^{\xi}_{\bf Q'} - \Ea^\eta_{\bf Q} \mp   \hbar\Omega_{j{\bf Q'-Q}} \right)$ as the phonon mediated scattering tensor.\\

The large twist angle in the experiment gives rise to very short  moire periods with a length scale comparable with the exciton Bohr radius. Therefore, a strong modification of the exciton center-of-mass motion, i.e. a moire-trapping of excitons is not expected \cite{brem2020tunable}. Therefore, we neglect the twist angle, aiming to a more qualitative description of the spatial exciton dependence. 
A spatial change in dynamics as observed in the experiment could be caused by several processes, such as spatially dependent dielectric environment or change in the layer distance. These changes have as common effect that the general energy landscape of the system is strongly modified. This can be  included in our model by introducing a spatially dependent relative energy shift $\Delta E_{\rm sp}$ between the two layer band structures. Note that we neglect possible changes in the tunnelling strength.


\bibliography{bibtexfile}

\end{document}